\documentclass[twocolumn,superscriptaddress,amsmath,amssymb]{revtex4}
\usepackage{graphicx}
\usepackage{dcolumn}
\usepackage{bm}
\usepackage{amsmath}
\begin{document}

\title{Quasi-particles ultrafastly releasing kink bosons to form Fermi arcs in a cuprate superconductor}

\author{Y.~Ishida}
\author{T.~Saitoh} 
\affiliation{ISSP, University of Tokyo, Kashiwa-no-ha, Kashiwa, Chiba 277-8581, Japan}

\author{T.~Mochiku}
\author{T.~Nakane}
\author{K.~Hirata}
\affiliation{National Institute for Materials Science, 1-2-1 Sengen, Tsukuba, Ibaraki 305-0047, Japan}

\author{S.~Shin}
\affiliation{ISSP, University of Tokyo, Kashiwa-no-ha, Kashiwa, Chiba 277-8581, Japan}
\affiliation{CREST JST, University of Tokyo, Kashiwa-no-ha, Kashiwa, Chiba 277-8581, Japan}

\date{\today}

\begin{abstract}
In a conventional framework, superconductivity is lost at a critical temperature ($T_c$) because, at higher temperatures, gluing bosons can no longer bind two electrons into a Cooper pair. In high-$T_c$ cuprates, it is still unknown how superconductivity vanishes at $T_c$. We provide evidence that the so-called $\lesssim$70-meV kink bosons that dress the quasi-particle excitations are playing a key role in the loss of superconductivity in a cuprate. We irradiated a 170-fs laser pulse on Bi$_2$Sr$_2$CaCu$_2$O$_{8+\delta}$ and monitored the responses of the superconducting gap and dressed quasi-particles by time- and angle-resolved photoemission spectroscopy. We observe an ultrafast loss of superconducting gap near the $d$-wave node, or light-induced Fermi arcs, which is accompanied by spectral broadenings and weight redistributions occurring within the kink binding energy. We discuss that the underlying mechanism of the spectral broadening that induce the Fermi arc is the undressing of quasi-particles from the kink bosons. The loss mechanism is beyond the conventional framework, and can accept the unconventional phenomena such as the signatures of Cooper pairs remaining at temperatures above $T_c$. 
\end{abstract}

\maketitle

The cuprate superconductivity is a novel state of matter that is believed to go beyond the theoretical description and concepts of superconductivity established in Bardeen-Cooper-Schrieffer (BCS)~\cite{BCS}. Understanding the unconventional superconductivity remains a major challenge in condensed-matter physics~\cite{Rev_Nature15}.

Angle-resolved photoemission spectroscopy (ARPES) is a powerful method to investigate the electronic structures of matter, and it has been continuously deepening the insights into the cuprates~\cite{RMP_ARPES,NPhys_ARPES}. Various peculiar electronic structures of the cuprates have been revealed by ARPES that can be summarized as follows: 
(1) The superconducting gap has a $d$-wave symmetry~\cite{RMP_ARPES};  (2) Across $ T_c$, the gap is diminished only near the $d$-wave node in momentum ($k$) space, resulting in a seemingly disconnected Fermi surface, or Fermi arcs, above $T_c$~\cite{Norman_Nature98}; (3) Quasi-particle dispersions commonly exhibit kink structures below $\sim$70~meV~\cite{Lanzara_kink}, indicating that individual-electron excitations are coherently dressed by some kink boson modes~\cite{Gormko,Devereaux_Cuk,Dahm_NPhys}. However, not only the origin of the kink bosons (phonons or some magnetic modes~\cite{Lanzara_kink,Gormko,Devereaux_Cuk,Dahm_NPhys,Sangiovanni}) but also their role, whether they are gluing the Cooper pairs or doing something else~\cite{Dessau_Broken}, has been a mystery. Moreover, the relationship, if at all, among the dressed quasi-particles, Fermi arcs, and $d$-wave superconducting gap remains unclear. It is also a challenge to understand the connection between the peculiar electronic structures and unconventional phenomena such as the signatures of incoherent Cooper pairs remaining at temperatures above $T_c$, which is an indication that the gluing of a Cooper pair is still substantial in the normal state~\cite{VanishingPhaseCoh,Nernst_Uchida,Nernst_YayuWang,Gomes_STM,NPhys_Josephson,JWLeeDavis_STM}.

Concerning the Fermi arc, several ARPES studies recently revealed a remnant feature of superconducting peaks existing in the Fermi-arc region in the normal state~\cite{Dessau_PRB,Dessau_NaturePhys12,Kondo_NCom15,Hashimoto_NMat14}. That is, the $d$-wave gap around the node is not closed when the Fermi arcs emerge at $T_c$~\cite{Kondo_NCom15}. What happens at $T_c$ is the broadening of the spectrum that fills the gap around the $d$-wave node~\cite{Dessau_PRB,Dessau_NaturePhys12,Kondo_NCom15}, a mechanism that is distinct from the conventional closure of the gap at $T_c$ in the BCS scheme. In addition, the remnant superconducting peaks in the Fermi-arc region were associated~\cite{Dessau_PRB, Dessau_NaturePhys12, Kondo_NCom15} to the signatures of incoherent Cooper pairs in the normal state~\cite{VanishingPhaseCoh,Nernst_Uchida,Nernst_YayuWang,Gomes_STM,NPhys_Josephson,JWLeeDavis_STM}. The next step would then be to understand the microscopic mechanism of the temperature-dependent spectral broadening that fills the $d$-wave gap around the node, forms the Fermi arcs, and quenches the macroscopic superconductivity at $T_c$.

Recently, a pump-and-probe method based on femto-second pulsed laser sources has been implemented into ARPES, and it became possible to investigate the responses of the cuprate's electronic structures to a light pulse~\cite{Perfetti_TTM,Bovensiepen,Lanzara_TrARPES_NPhys,Lanzara_Science,Lanzara_NCom,Bovensiepen_PRB14,Smallwood_PRB14,Shen_PRL15,Perfetti_PRB15}. In time-resolved ARPES (TrARPES), a pump pulse is impinged on a sample, and a probe pulse snapshots the non-equilibrated state at a certain delay time $t$ from the impact (Fig.~1a). Direct observation of the responses of quasi-particles along the $d$-wave node opened pathways to investigate electron-boson couplings~\cite{Perfetti_TTM,Shen_PRL15}, self-energy correlations due to kink bosons~\cite{Lanzara_TrARPES_NPhys,Lanzara_NCom,Smallwood_PRB14}, photo-dopings~\cite{Bovensiepen_PRB14}, and Cooper-pair coherence~\cite{Perfetti_PRB15}. Studies over a wide $k$ region showed carrier and gap dynamics depending on pump fluence and equilibrium temperature~\cite{Bovensiepen,Lanzara_Science,Smallwood_PRB14}. We here report on the investigation of Bi$_2$Sr$_2$CaCu$_2$O$_{8+\delta}$ (Bi2212) by TrARPES. We succeed in visualizing the pump-induced emergence of Fermi arcs, and find that the emergence is accompanied by 
the increase of quasi-particle scattering rate characterized by spectral broadenings and weight redistribution within the kink binding energy. 
Based on the results, we discus  the hitherto unnoticed role of the kink bosons in the spectral broadenings relevant to the unconventional loss of superconductivity at $T_c$.

\begin{figure*}[htb]
\begin{center}
\includegraphics[width=14cm]{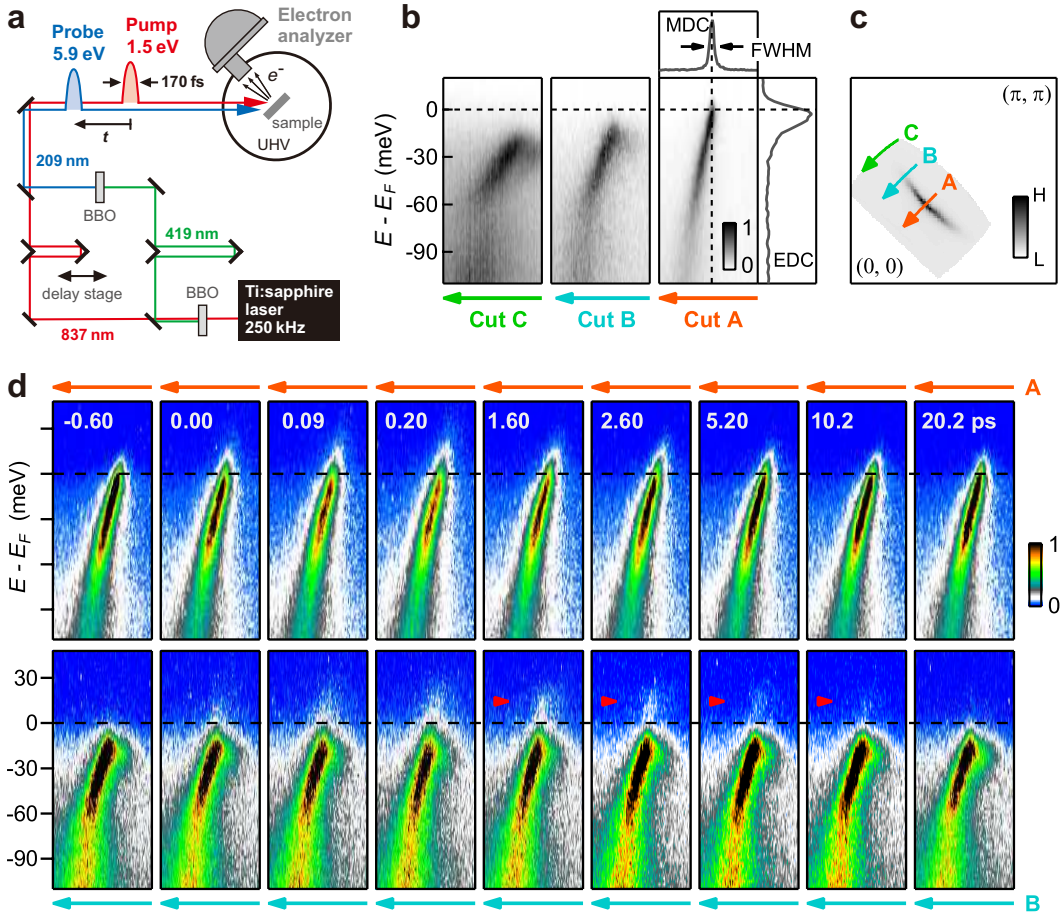}
\caption{\label{fig1} {\bf TrARPES of Bi2212.} 
(a) Schematic of the pump-and-probe experiment. The interval of the pump-and-probe pulses arriving on the sample is controlled by a mechanical delay stage. (b) Quasi-particle dispersions recorded by the 5.9-eV probe at 10~K along cuts A-C in $k$ space. Across $E_F$, a gap is opened that increases in going away from the nodal (0,\,0)\,-($\pi$,\,$\pi$) (cut A) to off-nodal directions (cuts B and C). (c) Remnant Fermi surface of Bi2212 represented by the spectral weight at $E_F\,\pm$\,5~meV mapped in $k$-space. Cuts A-C are also shown by arrows. (d) TrARPES images recorded at 10~K along cuts A and B shown in the top and bottom panels, respectively. Arrows in the bottom panels indicate the pump-induced population of the superconducting peak in the unoccupied side. 
}
\end{center}
\end{figure*}

\subsection*{Results}

In order to detect the response of the $d$-wave gap that becomes vanishingly small on approach to the node in $k$ space, we achieved the 10.5-meV energy resolution by adopting a fairly monochromatic pulses of 170-fs duration and reducing the space-charge broadening effect~\cite{Ishida_RSI}.  With this setup, the characteristic electronic structures of Bi2212 ($T_c$\,$\sim$\,92~K)~\cite{Mochiku} were nicely observed by the 5.9-eV probe~\cite{Ishida_RSI}: Across the Fermi level ($E_F$), the quasi-particle dispersions exhibit the $d$-wave gap (Fig.~1b); The dispersion kinks are observed around -70~meV  (Fig.~1b);  Spectral weight is distributed in $k$ space along the remnant Fermi surface (Fig.~1c).

\begin{figure}[htb]
\begin{center}
\includegraphics[width=8.3cm]{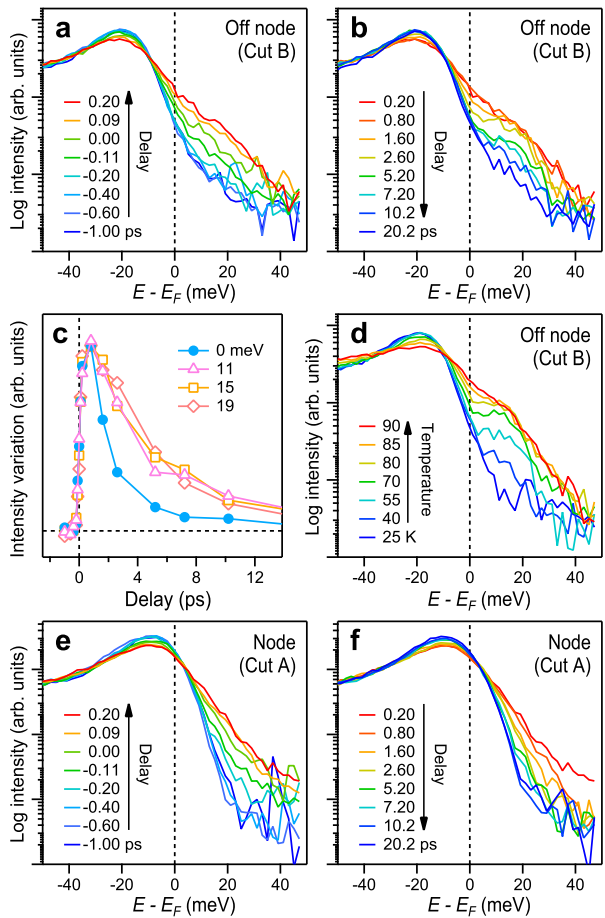}
\caption{\label{fig2} {\bf Collapse-and-recovery of the superconducting gap.} 
(a and b) EDCs at $k$-space where cut B crosses the remnant Fermi surface; see Fig.~1c and Ref.~\cite{Ishida_RSI}. Spectra recorded at $t$\,$\le$\,0.2 (a) and at $t$\,$\ge$0.2~ps (b) are displayed. (c) Variation of the spectral intensity of the EDCs in the superconducting gap ($E\,-\,E_F$\,=\,0~meV) and in the superconducting peak region above $E_F$ (11, 15, and 19~meV). Spectral intensity is taken in the energy window of $\pm$3~meV and the curves are normalized to the area around 0~ps. (d) Temperature dependence of EDCs, which were recorded without irradiating the pump. (e and f) EDCs at the node recorded at $t$\,$\le$\,0.2 (e) and at $t$\,$\ge$0.2~ps (f). 
}
\end{center}
\end{figure}

TrARPES images recorded at various $t$'s at the pump fluence of 14~$\mu$J/cm$^2$ are shown in Fig.~1d for the nodal (cut A, upper panels) and off-nodal (cut B, lower panels) directions. On arrival of the pump, the spectral intensity is spread into the unoccupied side ($E - E_F$\,$>$\,0) for both cuts A and B. As a result, the superconducting gap at cut B is smeared out; see the images recorded at $t$\,=\,0.09 and 0.20~ps. Subsequent recovery dynamics lasts over 10~ps, during which, some spectral weight occurs around 15~meV as marked by arrows in the lower panels. This is attributed to the pump-induced population of the superconducting peak in the unoccupied side developed across the superconducting gap. The superconducting gap is thus recovering after $\sim$0.2~ps.

More details of the collapse-and-recovery of the superconducting gap can be obtained by analyzing the energy distribution curves (EDCs). On arrival of the pump (Fig.~2a), the EDC at cut B tails into the unoccupied side and the superconducting gap is filled. During the recovery (Fig.~2b), a shoulder is formed around 15~meV, because the intensity around $E_F$ decreases faster than that around 15~meV (Fig.~2c). This demonstrates the recovery of the superconducting gap and transient population of the superconducting peak above $E_F$. Thermally-populated superconducting peak indeed occurs at $\sim$15~meV recorded at elevated temperatures (Fig.~2d). The pump-induced spread of the spectral intensity into the unoccupied side at $t$\,=\,0.20~ps (Fig.~2b) is slightly exceeding that induced thermally at 90~K (Fig.~2d). Therefore, the effective electronic temperature $T_e$ just after the pump at the fluence of 14~$\mu$J/cm$^2$ is slightly exceeding $T_c$, which is in good agreement with the previous studies showing similar critical fluence~\cite{Smallwood_PRB14}. The EDCs at the node are shown in Figs.~2e and 2f. The spectral shape around $E_F$ mostly obeys that of a thermal distribution function, which is consistent with the standpoint of the previous TrARPES study~\cite{Perfetti_TTM}. The main role of the pump can thus be regarded as to elevate the temperature of the electronic system. Nevertheless, although weak in intensity, we also discern a plateau extending into the unoccupied side in the spectra recorded at $t$\,$\le$\,0.20~ps. A similar plateau is observed in the spectra of graphite upon pumping~\cite{Ishida_HOPG}. This indicates that a small portion of the excited electrons are non-thermally distributed during the irradiation of the pump pulse of 170~fs in width.

\begin{figure}[h]
\begin{center}
\includegraphics[width=7.9cm]{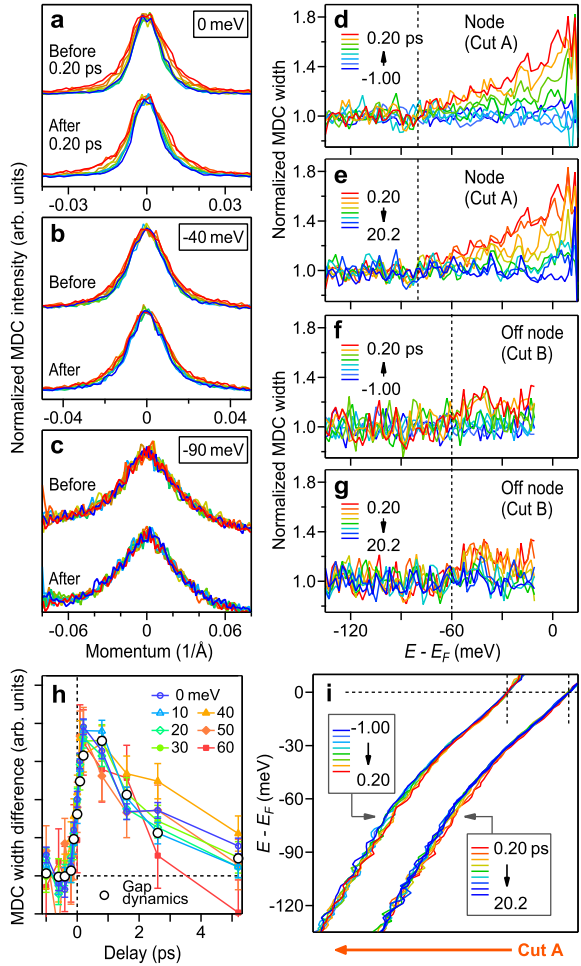}
\caption{\label{fig3} {\bf Ultrafast response of the quasi-particles.} 
(a\,-\,c) MDCs at 0 (a), -40 (b), and -90~meV (c) of the nodal quasi-particles. Here, the MDCs were normalized to the peak height and referenced to the peak position in order to show clearly the changes in their width. The MDCs recorded at $t$\,$\ge$\,0.20~ps are shifted vertically. (d - g) Spectra of MDC width of the nodal (d and e) and off-nodal (f and g) cuts (cuts A and B, respectively, of Fig.~1c) normalized to those before pumping.  Vertical dashed lines indicate the energy above which pump-induced variations become pronounced. (h) Pump-induced variations in the MDC width at various energies. The curves are normalized to the area around 0~ps. The line profile representing the gap dynamics (see, text) is also overlaid. (i) Pump-induced changes in the nodal quasi-particle dispersion. The dispersions recorded at $t$\,$\ge$\,0.20~ps are shifted horizontally. The dispersions are represented by the MDC peak positions. The color codings in (a\,-\,g and i) follow those adopted in Figs.~1a and 1b. 
}
\end{center}
\end{figure}

We investigate below how the dispersions and scattering rates of the quasi-particles respond during the collapse-and-recovery of the superconducting gap. This is done by analyzing the momentum distribution curves (MDCs) of the quasi-particle spectrum.

\begin{figure*}[htb]
\begin{center}
\includegraphics[width=16cm]{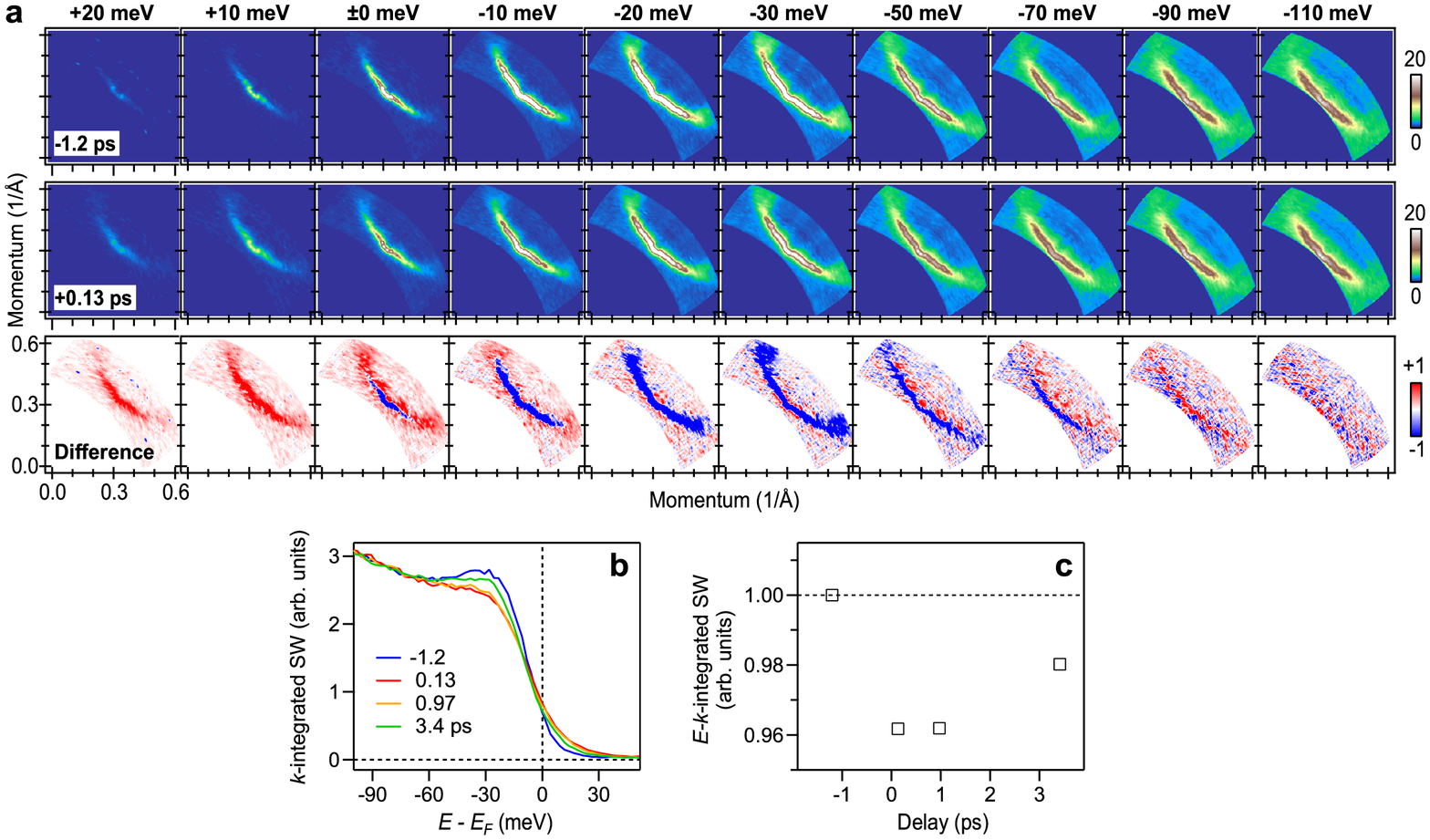}
\caption{\label{fig4} {\bf Spectral-weight variation in $E$\,-\,$k$ space upon the collapse of the superconducting gap.} 
(a) Spectral weight maps at various energies (integral window of $\pm$\,5~meV) recorded at -1.2~ps (top) and at 0.13~ps (middle). The bottom row shows the maps of difference between those recorded at 0.13 and -1.2~ps. The maps are symmetrized along the nodal direction. (b) Pump-induced variation of sDOS. (c) Intensity of sDOS at [-80, 50~meV] as a function of delay showing spectral-weight conservation within 5~\%. 
}
\end{center}
\end{figure*}

Figures~3a\,-\,3c show MDCs along the nodal cut A at several energies. Here, the peak positions and heights are aligned to highlight the changes in the MDC width representing the quasi-particle scattering rate. At $E_F$ (Fig.~3a), the MDC shows pump-induced broadening and subsequent recovery over 10~ps, while such variations are less prominent at -40~meV (Fig.~3b); At -90~meV (Fig.~3c), the variations are negligibly small. The spectra of the variation ratio of the MDC width (MDC width divided by that before the pump) for the nodal cut A are shown in Figs.~3d and 3e, while those for the off-nodal cut B are shown in Figs.~3f and 3g. The spectra for cut B are truncated around -10~meV because the dispersion around $E_F$ was gapped at $t  <$ 0. Clearly, the variations are seen only within the 70-meV binding energy for the nodal cut A (Figs.~3d and 3e). These variations are similar to those reported in the ARPES study by Kaminski {\it et al.}~\cite{Kaminski_TDep_Dispersion}: MDC broadened only within the kink binding energy when the equilibrium temperature was increased across $T_c$~\cite{Kaminski_TDep_Dispersion}. That is, the variations seen in TrAPRES are nicely tracking the temperature-dependent changes seen in ARPES. For the off-nodal cut B (Figs.~3f and 3g), the pump-induced variations occur in a shallower energy region than those occurring in the nodal cut A. The pump-induced spectral broadening confined in a shallower energy region in going away from the nodal direction nicely corresponds to the momentum dependence of the kink binding energy being largest in the nodal direction~\cite{Devereaux_Cuk}.

The MDC width changing only within the kink binding energy indicates that the pump decouples the quasi-particle excitations from the kink boson modes of $\lesssim$70~meV. That is, the low energy excitations within the kink binding energy experience increased scatterings by the boson modes that are no more coherently dressed in the pump-induced state. The transient change of the nodal MDC width for representative energies at $>$-70 meV are shown in Fig.~3h. We observe that the line profiles are mostly independent of energy, indicating that the modification of the quasi-particle scattering rate is governed by a single parameter such as the number of the scatterer. In Fig.~3h, we also overlaid the spectral weight in the gap at cut B (reproduced from Fig.~2c), which nicely represents the dynamics of the gap~\cite{Kondo_NPhys}. Both profiles matched nicely, suggesting that there is a link between the gap filling and spectral broadening, as we discuss later. The nodal quasi-particle scattering rate affected by forward scatterings may also increase when $T_e$ is elevated~\cite{ElasticInelastic}, but these scattering channels also alter the scattering rates beyond the kink binding energy. The robustness of the MDC width beyond the kink, which is also observed in the temperature-dependent ARPES study~\cite{Kaminski_TDep_Dispersion}, therefore excludes the dominance of the pump-induced variation in the forward-scattering channels.

The response of the nodal quasi-particle dispersion is presented in Fig.~3i. We observe that the pump-induced changes are similar to those observed when temperature is increased across $T_c$~\cite{Lanzara_kink,Kaminski_TDep_Dispersion,WSLee_TDep_Dispersion,Zhou_TDep_Dispersion}: Both the pump and equilibrium heating induce changes particularly around the -70-meV kink region but not around $E_F$. Therefore, concerning the response dependency of the dispersion kinks as well as the MDC widths, the 170-fs pulse has an effect similar to increasing $T_e$. The undressing of the quasi-particles from the spectrum of the kink bosons at $<$70~meV is thus occurring upon the increase of temperature. We here note that the electron-boson coupling constant need not change for the decoupling to occur upon heating; Fore example, see the temperature dependence of the electron self energy in an Einstein model~\cite{PhononContribution,Marsiglio}, although considerations are required in cuprates such as the renormalization of the dispersion kinks due to anti-ferromagnetic fluctuations~\cite{Sangiovanni}, momentum dependence of the couplings~\cite{Devereaux_Cuk}, and spectrum of the modes existing at $\lesssim$70~meV~\cite{WSLee_TDep_Dispersion}. 
Non-thermal electron distributions and the so-called phonon-window effect~\cite{PRX13} may be realized at shorter time scales around $t\,=\,0$ and/or at low-temperature and low-pump-fluence limit as discussed in Ref.~\cite{Shen_PRL15} 
(Also see Refs.~\cite{PRB94_Aeschlimann,Calandra-Mauri} describing the discrepancy between the lifetimes of a single-particle excitation and an excited ensemble of electrons).

Finally, we present the pump-induced Fermi arc and investigate how the spectral weight is varied in $E$\,-\,$k$ space when the superconducting gap is collapsed. Figure~4a shows the spectral weight mapped in $k$ space for various energies. The top and middle rows respectively show the maps recorded before the pump ($t$\,=\,-1.2~ps) and upon the collapse of the superconducting gap ($t$\,=\,0.13~ps) at the fluence of 14~$\mu$J/cm$^2$; the bottom row shows their difference. Here, we took advantage of TrARPES that the spectral intensity sampled at different $t$'s can be compared directly without any data processing, because the intensities in the raw data are already normalized to the acquisition time; See, Methods and Ref.~\cite{Ishida_RSI}. At $E_F$, the spectral weight is localized around the node before the pump (top row). Upon the collapse of the superconducting gap (middle row), the spectral-weight distribution at $E_F$ is elongated towards the off node. This is the evolution of the Fermi arc~\cite{Dessau_NaturePhys12} induced by the pump. At -30~meV, strongly spectral-weight depleted regions, or hot spots, occur away from the node, which highlight the termination of the arc in the difference image. The region beyond the arc is dominated by pseudo-gaps~\cite{NPhys_ARPES}. At $\leq$-90~meV, there is hardly any change in the spectral weight, which is in accord with the robustness of the MDCs beyond the kink. Overall, the spectral weight is varied within the $\sim$70-meV kink binding energy upon the collapse of the superconducting gap; Seen at $E_F$, the Fermi arc evolves.

Figure~4b displays the spectral density of states (sDOS), which is the spectral weight integrated over the $k$ space surveyed. Pump-induced changes in the sDOS occur at $\gtrsim$-70~meV with the spectral weight at [-80,~50~meV] being conserved within the accuracy of 5~\%, as shown in Fig.~4c. Thus, the spectral variation induced by the pump is characterized by the spectral-weight redistribution within the kink binding energy: The redistributed spectral weight within the kink binding energy is smearing out the gap around the node.

\subsection*{Discussion}

The TrARPES results showed that the following two are not separate entities: The quasi-particle undressing from the kink boson modes and related increase in the quasi-particle scattering rate; collapse of the gap near the node and emergence of the Fermi arcs. The results indicate a mechanism by which the undressing of quasi-particles from the kink boson modes drives the spectral broadening that causes the unconventional gap filling near the node, emergence of Fermi arcs, and loss of macroscopic superconductivity. The existence of the node is important for the loss mechanism to function, because it is the small gap around the node that is prone to the temperature-dependent spectral broadening caused by the undressing. The role of the kink boson modes herein is apparently not to bind a Cooper pair. In the conventional theory~\cite{BCS}, the loss of superconductivity at $T_c$ is due to the shrink of the Cooper-pair binding energy 2$\varDelta$. Concepts of quasi-particle undressing from the kink boson modes are not included therein. If the undressing upon heating can be suppressed, for example, by strengthening the electron-boson coupling constant, $T_c$ may increase. Since $\varDelta$ can be non-zero even above $T_c$, our picture accepts the existence of incoherent Cooper pairs above $T_c$~\cite{VanishingPhaseCoh,Nernst_Uchida,Nernst_YayuWang,Gomes_STM,NPhys_Josephson,JWLeeDavis_STM}, whose gluing bosons may likely differ from the kink boson modes.

\subsection*{Methods}
Slightly under-doped Bi2212 single crystals ($T_c\sim$\,92~K) were grown by the traveling solvent floating zone method~\cite{Mochiku}.
Samples were cleaved in the TrARPES spectrometer at the base pressure of $<$5\,$\times$10$^{-11}$~Torr. In order to achieve a high energy resolution in TrAPRES, we adopted a Ti:Sapphire laser system that delivered 1.5-eV pulses of 170-fs duration, which was not too short~\cite{Ishida_RSI}. Note, due to the uncertainty principle, monochromaticity and short duration of a pulse are incompatible. The portion of the laser pulse was up-converted into 5.9-eV probe by using two non-linear crystals, BBO; see Fig.~1a. During the measurements, we reduced the probe intensity until the space-charge-induced shift of the spectrum was below the detection limit ($\ll$1~meV); a too intense probing pulse generates a bunch of photoelectrons that repel each other through the Coulomb effect, thereby degrading the energy-and-angular resolution in the spectra. The photoelectrons were collected by a VG Scienta analyzer at the energy resolution of 10.5~meV. The spacial overlap of the pump and probe beams, origin of the delay, and time resolution of 350~fs at the sample position were calibrated by using pump-and-probe photoemission signal of graphite attached next to the sample. TrARPES measurements were carried out at the pump intensity of $\sim$14~$\mu$J/cm$^{2}$ and cryostat temperature of $T\,=\,10$~K. The pump-beam induced heating effect was +10~K at most, judged from the broadening of the Fermi cutoff at the node recorded at $t$\,$<$\,0. The set of TrARPES images was acquired by repetitively scanning the delay stage until a sufficient signal-to-noise ratio was achieved. Such a dataset acquisition allowed us to evaluate the pump-induced variation of the spectral weight without normalizing the intensity of the spectra recorded at different delay time values. In principle, spectral-weight transfer with temperature can also be investigated by recording ARPES images at various temperatures, although normalization-free analysis is difficult because various factors have to be concerned such as the change in the surface condition and shift of the sample position due to thermal expansion of the cryostat. See Ref.~\cite{Ishida_RSI} for the details of the energy resolution, reduction of the space-charge broadening effect, energy calibration of the Bi2212 spectra, determination of the time origin, and advantage of the repetitive delay-stage scanning during the data acquisition. 
\\

\noindent
\textbf{Acknowledgements:}

\noindent
This work was supported by Photon and Quantum Basic Research Coordinated Development Program from MEXT and JSPS KAKENHI, Grant Nos.~23740256 and 26800165. 
\\

\noindent
\textbf{Author contributions:}

\noindent
T.M., T.N., and K.H.~synthesized and characterized the samples. Y.I.~and T.S.~performed the experiments and analyzed the data. Y.I.~wrote the manuscript. S.S.~supervised the project. All the authors commented on the manuscript and contributed to discussion. 
\\

\noindent
\textbf{Competing Financial Interests:}

\noindent
The authors declare no competing financial interests. 
\\

\end{document}